# Quantum and thermal spin relaxation in diluted spin ice: $Dy_{2-x}M_xTi_2O_7$ (M = Lu, Y)

J. Snyder[1], B. G. Ueland[1], Ari Mizel[1], J. S. Slusky[2], H. Karunadasa[2], R. J. Cava[2], and P. Schiffer[1*]

[1]*Department of Physics and Materials Research Institute, Pennsylvania State University, University Park PA 16802*

[2]*Department of Chemistry and Princeton Materials Institute, Princeton University, Princeton, NJ 08540*

**Abstract**

We have studied the low temperature a.c. magnetic susceptibility of the diluted spin ice compound $Dy_{2-x}M_xTi_2O_7$, where the magnetic Dy ions on the frustrated pyrochlore lattice have been replaced with non-magnetic ions, M = Y or Lu. We examine a broad range of dilutions, $0 \leq x \leq 1.98$, and we find that the $T \sim 16$ K freezing is suppressed for low levels of dilution but re-emerges for $x > 0.4$ and persists to $x = 1.98$. This behavior can be understood as a non-monotonic dependence of the quantum spin relaxation time with dilution. The results suggest that the observed spin freezing is fundamentally a single spin process which is affected by the local environment, rather than the development of spin-spin correlations as earlier data suggested.

[*]*schiffer@phys.psu.edu*

## I. Introduction

Geometrically frustrated magnetic materials, in which the lattice geometry leads to frustration of the spin-spin interactions, display a variety of novel magnetic behavior [1]. Of particular recent interest are the rare earth pyrochlores spin ice materials such as $Dy_2Ti_2O_7$, $Ho_2Ti_2O_7$, and $Ho_2Sn_2O_7$ [2,3,4,5,6,7,8,9,10,11,12,13,14,15,16]. In these materials the lattice geometry and spin symmetry lead to frustration of ferromagnetic and dipolar interactions [17,18,19,20] and an unusual disordered ground state that has been demonstrated experimentally through neutron scattering studies [3,10,14] and also through measurements of the magnetic specific heat [4,12] and the a.c. magnetic susceptibility [16]. While the spin entropy only freezes out below $T_{ice}$ ~ 4 K in $Dy_2Ti_2O_7$, a.c. magnetic susceptibility studies show a strongly frequency dependent spin-freezing at $T_f$ ~ 16 K [5,7], below which the high frequency susceptibility ($f$ > 100 Hz) is suppressed. Because of the high degree of structural and chemical order in this material, the Dy spins respond to external fields with a very narrow distribution of spin relaxation times around a single characteristic relaxation time, $\tau$. While $\tau(T)$ is thermally activated for $T > T_{cross}$ (where $T_{cross}$ ~ 13 K), below $T_{cross}$, $\tau(T)$ exhibits relatively weak temperature dependence due to a crossover from thermal to quantum spin relaxation [6,15,16]. Neutron spin-echo studies suggest that all of the spin dynamics for $T > T_{ice}$ are associated with single spin processes [6], while dilution studies over a limited range of dilution [5] suggest that spin-spin correlations are important even at these high temperatures (which are well above the energy scale of the spin-spin interactions).

To further study the spin dynamics of the spin ice system, we have diluted the magnetic sublattice by replacing magnetic $Dy^{3+}$ ions with non-magnetic $Y^{3+}$ and $Lu^{3+}$ ions. In doing so, we extended previous studies of diluted spin ice [5] to a much broader range of dilutions, examining $Dy_{2-x}R_xTi_2O_7$ for $0 \leq x \leq 1.98$ (i.e. between a complete magnetic lattice to 1% filling of the lattice with magnetic $Dy^{3+}$ ions). Dilution studies of other geometrically frustrated magnets have lent considerable insight into the nature of the low temperature behavior [21,22,23,24]. For example, $SrCr_{8-x}Ga_{4+x}O_{19}$ (SCGO), a frustrated antiferromagnet with a layered kagomé structure [25,26] has been studied extensively as a function of the concentration of the Cr ions relative to non-magnetic Ga. While not possessing the level of disorder traditionally associated with spin glasses, SCGO displays a spin glass transition at low temperatures ($T_f \sim 5$ K << $\Theta_{Weiss} \sim 500$ K), which is associated with the geometrical frustration inherent to the kagomé lattice. Dilution is expected to have a significant impact on such a strongly frustrated system, and indeed $T_f$ in SCGO is strongly reduced by dilution [27,28], decreasing by ~ 32% for $x = 2$ (~25% dilution).

In the present study, we use a.c. magnetic susceptibility to probe the nature of the Dy spin relaxation as a function of dilution in both the thermal and the quantum spin relaxation regimes. We find that the $T_f \sim 16$ K spin-freezing is suppressed for low levels of dilution but re-emerges for $x > 0.4$ and persists to $x = 1.98$. This behavior is reflected in the spin relaxation time, which appears to increase monotonically with dilution in the thermally activated regime while it has distinctly non-monotonic behavior in the quantum relaxation regime.

## II. Experimental Results

Polycrystalline powder samples of $Dy_{2-x}M_xTi_2O_7$ samples were prepared using standard solid-state synthesis techniques described previously [5,29]. We used non-magnetic ions, M = Y and Lu, which have nearly the same ionic radius as Dy. X-ray diffraction demonstrated the samples to be single-phase with the pyrochlore structure. We studied the magnetization ($M$) as well as the real and imaginary parts ($\chi'$ and $\chi''$) of the a.c. susceptibility ($\chi_{ac}$) using a Quantum Design MPMS SQUID magnetometer and the ACMS option of the Quantum Design PPMS cryostat respectively.

Curie-Weiss fits to the high temperature d.c. magnetization yielded the effective moment per ion for all samples studied to be consistent with $J$ = 15/2 $Dy^{3+}$ ions. To confirm that dilution did not alter the single-ion anisotropy which is essential to the spin ice ground state, we measured the saturation magnetization. In the pure sample, the crystal field induced anisotropy reduces the saturation moment to be half the free ion value [10], which would presumably be changed systematically with dilution if the anisotropy were altered by the dilution. As can be seen in Figure 1, the saturation moment is essentially unchanged with dilution, demonstrating that dilution does not measurably alter the anisotropy of the system [10].

Figures 2-4 show the measured temperature dependent a.c. susceptibility of samples across our range of dilution with both Y and Lu at a characteristic frequency of 1 kHz. The undiluted sample shows the previously observed freezing transition, as indicated by a drop in $\chi'$ ($T$) and a corresponding rise in $\chi''(T)$ at $T_f$ ~ 16 K [5,7] ($T_f$ depends on the frequency of measurement, as discussed below). The drop in $\chi'(T)$ indicates that the spins' dynamic response is slowed such that they cannot respond to the time-varying magnetic field for $T < T_f$. This implies that the system is out of equilibrium on the time scale of the measurement, i.e. that the spin relaxation time is longer than the inverse of the frequency of the a.c. measurement. The appearance of this freezing temperature is sharply suppressed with dilution of the Dy up to $x = 0.4$ as we have reported previously [5], suggesting that the freezing is a cooperative effect. Our new data, which go to higher levels of dilution, surprisingly show that the freezing transition is reentrant in dilution, i.e. rather than being further suppressed with increasing dilution for $x > 0.4$, it becomes strikingly more pronounced as the dilution is increased. Furthermore, the freezing temperature increases with higher dilution -- up to $T_f \sim 21$ K for $x = 1.98$ at this frequency. This reentrance of the freezing transition is seen for both Lu and Y dilution as shown in figures 2 and 3, although there are differences between samples diluted with the two different ions, as shown in figure 4.

We can also characterize the effects of dilution by fitting the frequency dependence of the freezing temperature to an Arrhenius law, $f = f_0 e^{-E_A/k_B T_f}$ where $E_A$ is an activation energy for spin fluctuations and $f_o$ is a measure of the microscopic limiting frequency in the system. We plot such data in figure 5, and we find that $E_A$ is of order the single ion anisotropy energy and $f_o$ is of order GHz, both physically reasonable numbers for individual spin flips. The fitted values of $E_A$ and $f_o$ are given in the inset to figure 5, and it is notable that $E_A$ increases substantially with increasing $x$, especially at larger values of $x$. Given the correspondence of $E_A$ to the single ion anisotropy, the dependence of $E_A$ on dilution suggests that the slight change in the lattice constants associated with substitution on the Dy site result in a change in the crystal field level spacing which effectively changes the magnitude of the single-ion anisotropy (although not its qualitative character). Indeed, x-ray diffraction studies of the Y diluted series indicates that the lattice parameter decreases monotonically with dilution which is consistent with the rise in $E_A$.

Another method of parameterizing the spin response to the time-dependent a.c. magnetic field is through the Casimir-du Pré relation [30] which predicts, for a single relaxation time $\tau$, that $\chi''(f) = f\tau\left(\frac{\chi_T - \chi_S}{1 + f^2\tau^2}\right)$ where $\chi_T$ is the isothermal susceptibility in the limit of low frequency and $\chi_S$ is the adiabatic susceptibility in the limit of high frequency. Data of this sort are shown in figure 6 for different levels of dilution in magnetic fields of 0, 5, and 10 kOe. We have previously demonstrated that measurements of the undiluted sample fit the above form at $T$ = 16 K fairly well with only a slight broadening at lower temperatures [5,15,16]. Note that this behavior is in sharp contrast to that in other dense magnetic systems exhibiting glass-like behavior, in which the peak typically spans several decades [31,32,33]. Since the $\chi''(f)$ data display clear maxima, we can use these data to obtain a characteristic spin relaxation time, $\tau$, where $1/\tau$ is the frequency of the maximum in $\chi''(f)$ at a given temperature. The changing peak position with temperature allows us to characterize the evolution of $\tau(T)$, and these data have previously been analyzed to demonstrate the crossover from thermal to quantum spin relaxation below $T_{cross}$ ~ 13 K in the undiluted sample [15,16].

The $\chi''(f)$ data show a single clear peak in zero field (figure 6a), but application of a field of 5 kOe results in a double-peak structure to $\chi''(f)$ near $T_{cross}$ (figures 6b and 7a) [15]. This double-peak apparently corresponds to the two relaxation mechanisms – a phenomenon which is enhanced by dilution, as can be seen in figure 7b. The double peaks are quite surprising since we expect only a single peak corresponding to whichever relaxation mechanism is faster (quantum or thermal). The application of a 5 kOe field apparently suppresses the quantum tunneling of some of the spins (depending on orientation) due to the increased splitting of the spin states in the field. These spins relax thermally instead, and the two relaxation mechanisms produce the double-peak structure in $\chi''(f)$. When the field is increased much higher as seen in figure 6c, the quantum relaxation is more strongly suppressed for most of the spins, since the final and initial states are farther separated in energy [15].

The evolution of the double-peak structure with decreasing temperature is shown in figure 7. While the two peaks are only slightly separated for the undiluted sample data (figure 7a), the $x$ = 0.4 sample data show two distinct peaks over a broad range of temperature (figure 7b). The temperature dependence of the peak positions for this diluted sample provide evidence that the two maxima in the data should be associated with quantum and thermal relaxation processes. The higher frequency peaks for the $x$ = 0.4 sample are at approximately the same frequency at each temperature in the figure, as expected for quantum spin relaxation. By contrast, the lower frequency peaks (indicated by the vertical arrows) clearly show a decrease in the peak frequency with decreasing temperature as expected for thermally induced spin relaxation. The difference between the diluted samples and the undiluted sample in the structure of $\chi''(f)$ is difficult to explain without a detailed analysis of the microscopic physics, since spin relaxation depends sensitively on the energy states of the Dy ions. Curiously, as seen in figure 7c, a quite similar enhancement of the double-peak structure is observed in powder samples of the undiluted compound which were potted in non-magnetic Stycast 1266 epoxy (the potted sample shows little difference from the loose powder sample in zero field data [16]). Since the epoxy is expected to have a considerably larger thermal contraction coefficient than the pyrochlore compounds, the epoxy potted material will be subjected to pressure at low temperature. The resultant changes in the data from the epoxy-potted material further suggest that some of the differences between the diluted and the pure samples are due to chemical pressure associated with dilution of the lattice with the non-magnetic ions. As in the case of the applied pressure from the epoxy's thermal contraction, this chemical pressure can affect both the phonon spectrum and the crystalline field which determines the tunneling barrier between spin states.

As mentioned above, we can use the maxima in $\chi''(f)$ to obtain a characteristic spin relaxation time, $\tau$, where $1/\tau$ is the frequency of the maximum in $\chi''(f)$ at a given temperature. In figure 8, we plot $\tau(T)$ for samples with a range of dilutions in magnetic fields of 0, 5, and 10 kOe. In cases where $\chi''(f)$ displayed two maxima, both are plotted in the figure, and points are not shown for temperatures where $\chi''(f)$ did not show a maximum in our frequency range. There are three clear temperature regimes to the data. At high temperatures, the relaxation time appears to be activated, as expected for a thermally driven process. Below $T \sim T_{cross}$, the temperature dependence becomes much weaker, as the primary relaxation mechanism becomes quantum mechanical. A strong temperature dependence re-emerges at the lowest temperatures, an effect which has been associated with the development of correlations among the spins [15,16]. One unusual feature in the undiluted sample data is the sharp decrease of $\tau(T)$ on cooling near $T_{cross}$ at 5 kOe, since one ordinarily would not expect the relaxation time to shorten as the temperature is lowered. This apparent decrease is an artifact of the way we obtain the values of $\tau$. Since we take $\tau$ from local maxima in $\chi''(f)$, at temperatures and fields where there is a crossover from thermal to quantum spin relaxation, there can be an apparent jump in $\tau(T)$ when the maximum switches between the thermal and quantum relaxation frequencies. The data from the diluted samples where both relaxation mechanisms yield maxima for some range of fields and temperatures confirm that the sharp decrease in $\tau(T)$ can be attributed to this crossover effect.

The data in figure 8 demonstrate that there is a non-trivial dependence of the relaxation time on dilution in this system, as could have been deduced from the non-monotonic dilution dependence of the spin freezing transition. This dependence is shown explicitly in figure 9, in which we plot $\tau(x)$ at temperatures of $T$ = 5, 11, and 16 K in zero magnetic field and in different fields at $T$ = 16 K. These data show that $\tau(x)$ rises almost monotonically with increasing non-magnetic dilution for $T$ = 16 K, which is above $T_{cross}$, and the rise is strictly monotonic for the non-zero magnetic fields. By contrast, for the two lower temperatures, which are below $T_{cross}$, $\tau(x)$ decreases initially with dilution of the full magnetic lattice, has a minimum near $x = 0.5$ (we note for reference that at x = 0.5 each tetrahedron has on average one corner replaced with a non-magnetic ion), and subsequently increases with increasing dilution up to the largest values of $x$ where $\tau$ remained within the frequency range of our measurement apparatus.

## III. Analysis and Discussion

The data presented above provide an improved understanding of the spin relaxation and the dynamic spin-freezing in $Dy_{2-x}M_xTi_2O_7$. Previously published data obtained by our group (for $x \leq 0.4$) showed the freezing transition disappearing with dilution and thus suggested that the transition was associated with spin-spin correlations [5]. The present data, showing the re-emergence of the freezing transition with increasing dilution and its presence up to 99% dilution (i.e. $x = 1.98$), imply instead that the freezing is fundamentally a single-ion phenomenon [6], akin to a superparamagnetic blocking transition (albeit a phenomenon affected by the spin-spin interactions as described below).

The explanation for this highly unusual re-entrance of the spin-freezing transition can be found in the data of figure 9, which shows the spin relaxation time initially decreasing with dilution (which would naturally suppress the spin freezing) and subsequently increasing (which would cause spin freezing to re-emerge). The non-monotonic nature of $\tau(x)$ can be understood as the combination of two competing effects. The energy barrier for spin flips increases with dilution, as is evidenced by both the increasing value of $E_A(x)$ we obtain from the Arrhenius law data of figure 5 and from the increasing value of $\tau(x)$ we observe at $T = 16$ K in the presence of a magnetic field. In both of these cases, the data are dominated by thermal processes, and thus the physical interpretation of a barrier increasing with dilution is unambiguous. This increasing barrier is presumably due to changes in the crystal field splitting associated with changes in the lattice constant resulting from the dilution. In fact, at lower temperatures where spin relaxation occurs through quantum tunneling, the large barrier of the very highly diluted samples makes $\tau(x)$ unobservably long (Figure 8). The increasing barrier with dilution explains the increase in $\tau(x)$ for larger $x$, even at lower temperatures where the spin relaxation is through quantum tunneling.

The increasing energy barrier for spin flips with dilution does not, however, explain the decrease at smaller values of $x$. As can be seen in the inset to figure 8, the quantum spin relaxation is sensitive to magnetic field. This is presumably due to the effects of aligning and separating the energy levels of different spins states through the Zeeman term in the spin Hamiltonian. As the magnetic lattice is diluted, the local fields felt by each Dy spin will presumably be reduced, thus increasing the degeneracy of the different spin states and consequently reducing the spin relaxation time. The existence of the minimum in $\tau(x)$ at $x \sim 0.4$ does not have an obvious physical significance other than corresponding to the point where there exists a crossover between the two competing effects on $\tau(x)$.

## IV. Conclusions

In summary, studies of the nearly full range of non-magnetic dilution of $Dy_2Ti_2O_7$ provide new evidence that the spin relaxation near the classical to quantum crossover temperature is dominated by single-ion effects. As the reentrance of the spin-freezing transition with dilution indicates, quantum spin relaxation processes can lead to other novel behavior in frustrated rare-earth magnets. The results imply that even such large spin systems, where the spins are typically treated classically, ought to be considered with the effects of quantum dynamics in mind. The results further point to the importance of dilution as a tool with which to probe these systems, since the dilution dependence clearly elucidates the complex nature of the spin relaxation processes in this important model system.

## ACKNOWLEDGEMENTS

We gratefully acknowledge helpful discussions with D. A. Huse, A. Kent, M. Gingras and support from the Army Research Office PECASE grant DAAD19-01-1-0021 and NSF grant DMR-0101318. A.M. gratefully acknowledges the support of the Packard Foundation.

**Figure captions**

**Figure 1.** Magnetization as a function of applied field for $Dy_{2-x}Y_xTi_2O_7$ with values of *x* from 0 to 1.98 at a temperature $T$ = 1.8 K. All samples show saturation at a value of approximately 5 $\mu_B$/Dy indicating that the crystalline field effects are not changed by dilution.

**Figure 2.** The real and imaginary parts of the a.c. susceptibility of $Dy_{2-x}Y_xTi_2O_7$ as a function of temperature at a frequency of 1 kHz. The freezing transition is initially suppressed and then enhanced with increasing dilution of the Dy sites with Y.

**Figure 3.** The real part of the a.c. susceptibility of $Dy_{2-x}Lu_xTi_2O_7$ as a function of temperature at a frequency of 1 kHz. The freezing transition is initially suppressed and then enhanced with increasing dilution of the Dy sites with Lu.

**Figure 4.** Comparison of the freezing transitions between Lu and Y doped samples at 1 kHz. While there are differences between the two, the trend with dilution is similar.

**Figure 5.** Arrhenius law fits of the freezing temperature dependence on frequency for a range of diluted samples of $Dy_{2-x}Y_xTi_2O_7$ and $Dy_{2-x}Lu_xTi_2O_7$. The values of the freezing temperature are obtained from the minimum in the slope of $\chi''$ which corresponds to the maximum in $\chi'$. The inset shows the resulting fit parameters as a function of *x*, indicating that the energy barrier for thermal relaxation increases substantially with dilution.

Squares show $E_A$ and circles show fo with open symbols for Y-diluted samples and filled symbols for Lu-diluted.

**Figure 6.** Frequency dependence of $\chi''$ for $Dy_{2-x}Y_xTi_2O_7$ in applied fields of 0, 5, and 10 kOe at 12 K. Note the double-peak nature of the data at 5 kOe which demonstrates the combination of thermal and quantum spin relaxation processes which are active at this temperature.

**Figure 7.** Frequency dependence of $\chi''$ in an applied field of 5 kOe, showing the evolution of the peak shape with decreasing temperature below $T_{cross}$. a. The undiluted sample shows a broadened peak for $T$ = 11 and 12 K indicating the presence of two underlying relaxation mechanisms. b. The sample with $x$ = 0.4 shows a clear double peak structure corresponding to the two relaxation mechanisms. c. An undiluted sample potted in epoxy shows much the same behavior as the diluted sample with two clear peaks.

**Figure 8.** The temperature dependence of the characteristic spin relaxation time, $\tau$, for $Dy_{2-x}Y_xTi_2O_7$ for a range of dilutions, $x$, in magnetic fields of 0, 5, and 10 kOe. The inset shows the magnetic field dependence of the relaxation time, $\tau(H)$, for the undiluted $Dy_2Ti_2O_7$ sample at $T$ = 6 K (reprinted from [15]).

**Figure 9.** The characteristic spin relaxation time as a function of dilution for $Dy_{2-x}Y_xTi_2O_7$ at $T$ = 5, 11, and 16 K in zero field, and at $H$ = 0, 5, and 10 kOe at $T$ = 16 K. A minimum is evident near $x$ = 0.4 for low temperatures and small fields.

Figure 1 Snyder *et al.*

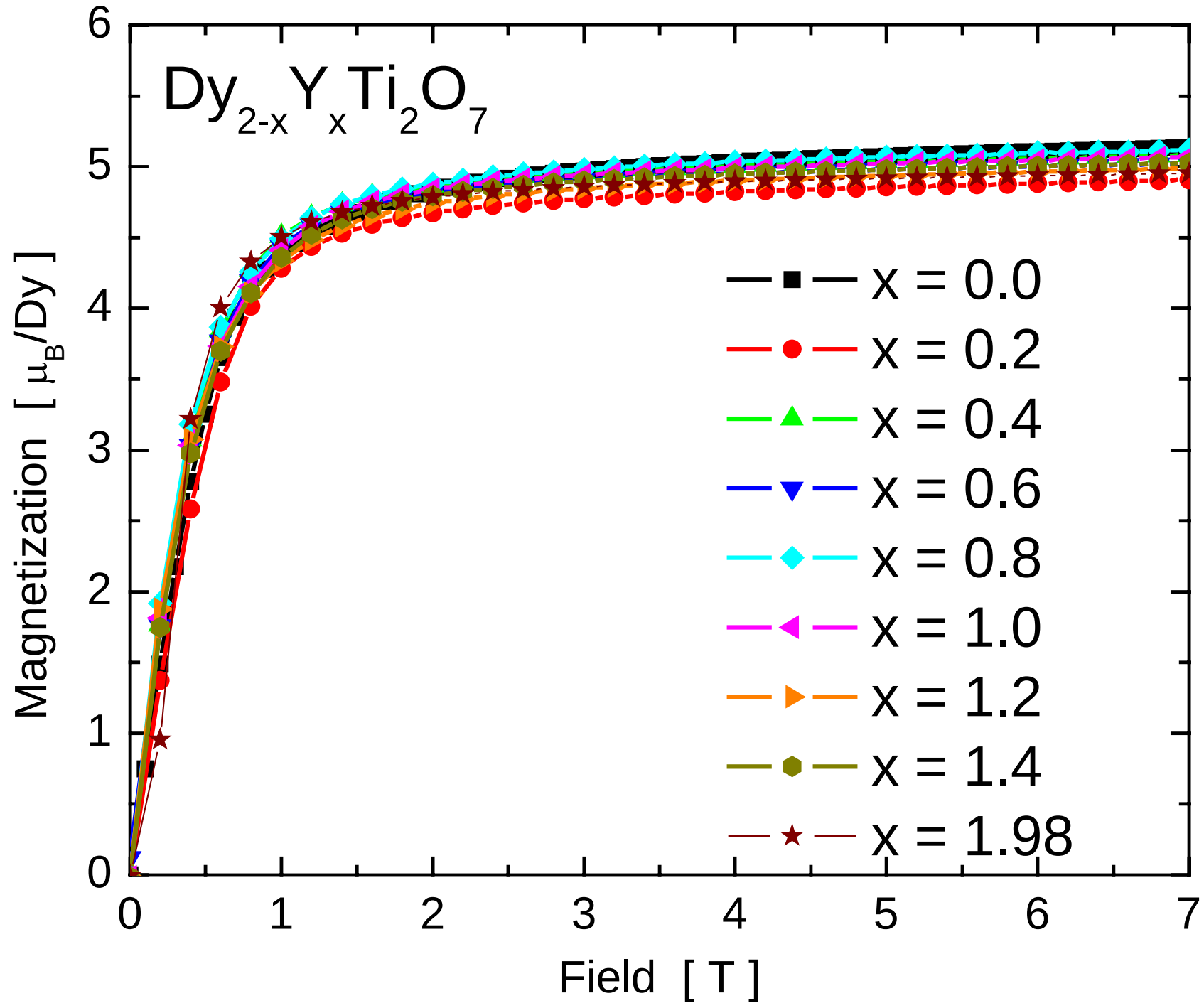

Figure 2 Snyder *et al.*

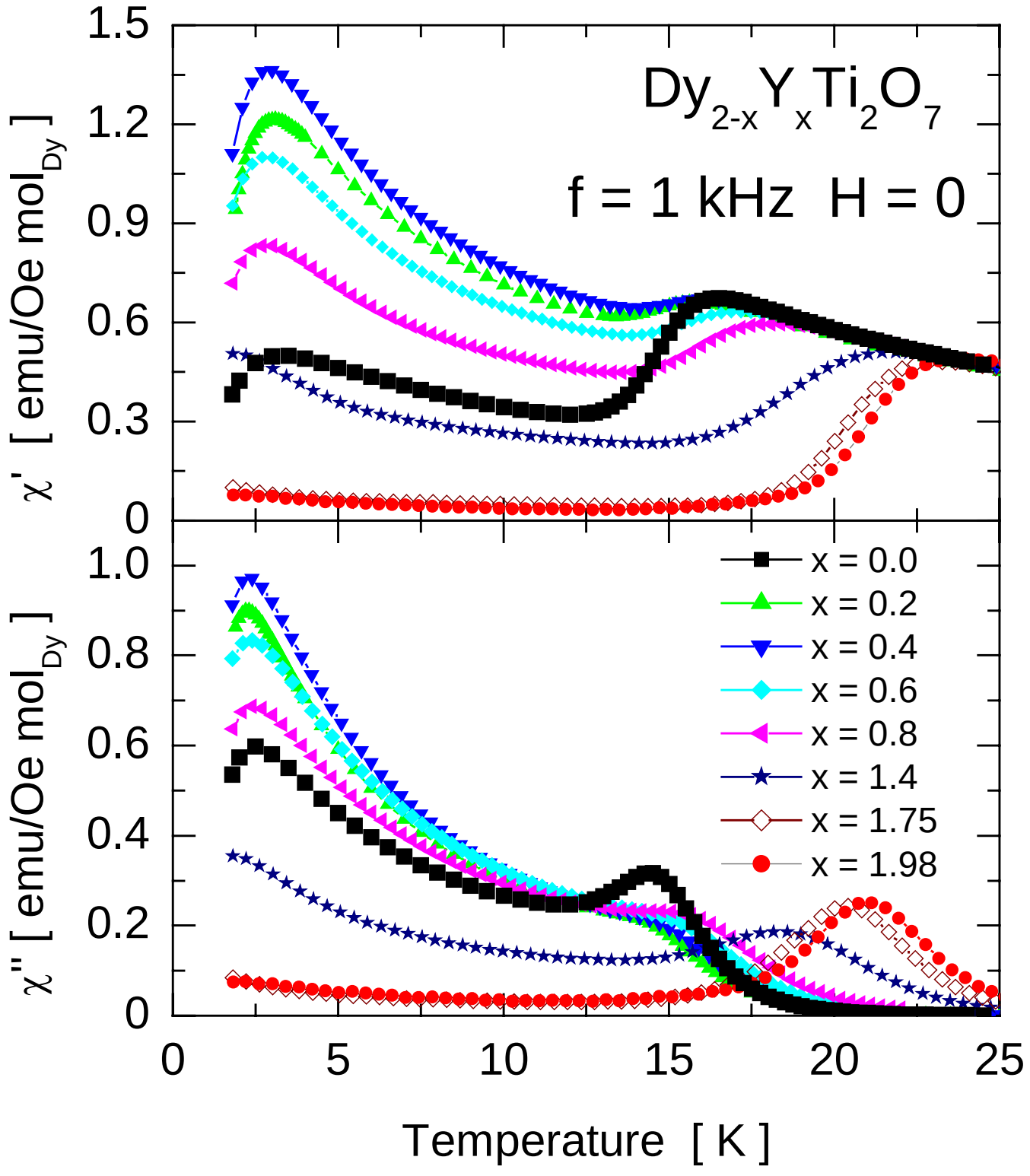

Figure 3    Snyder *et al.*

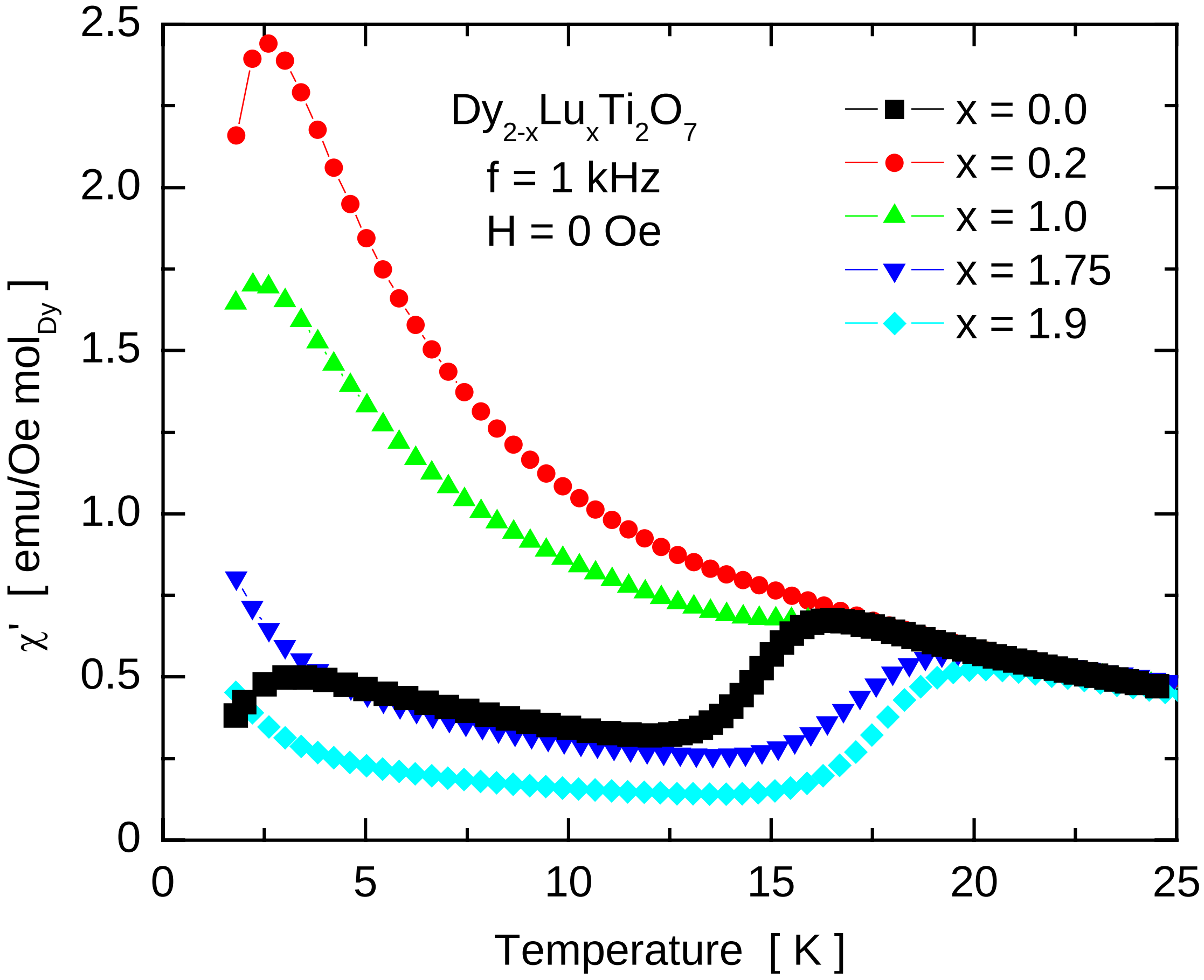

Figure 4 Snyder *et al.*

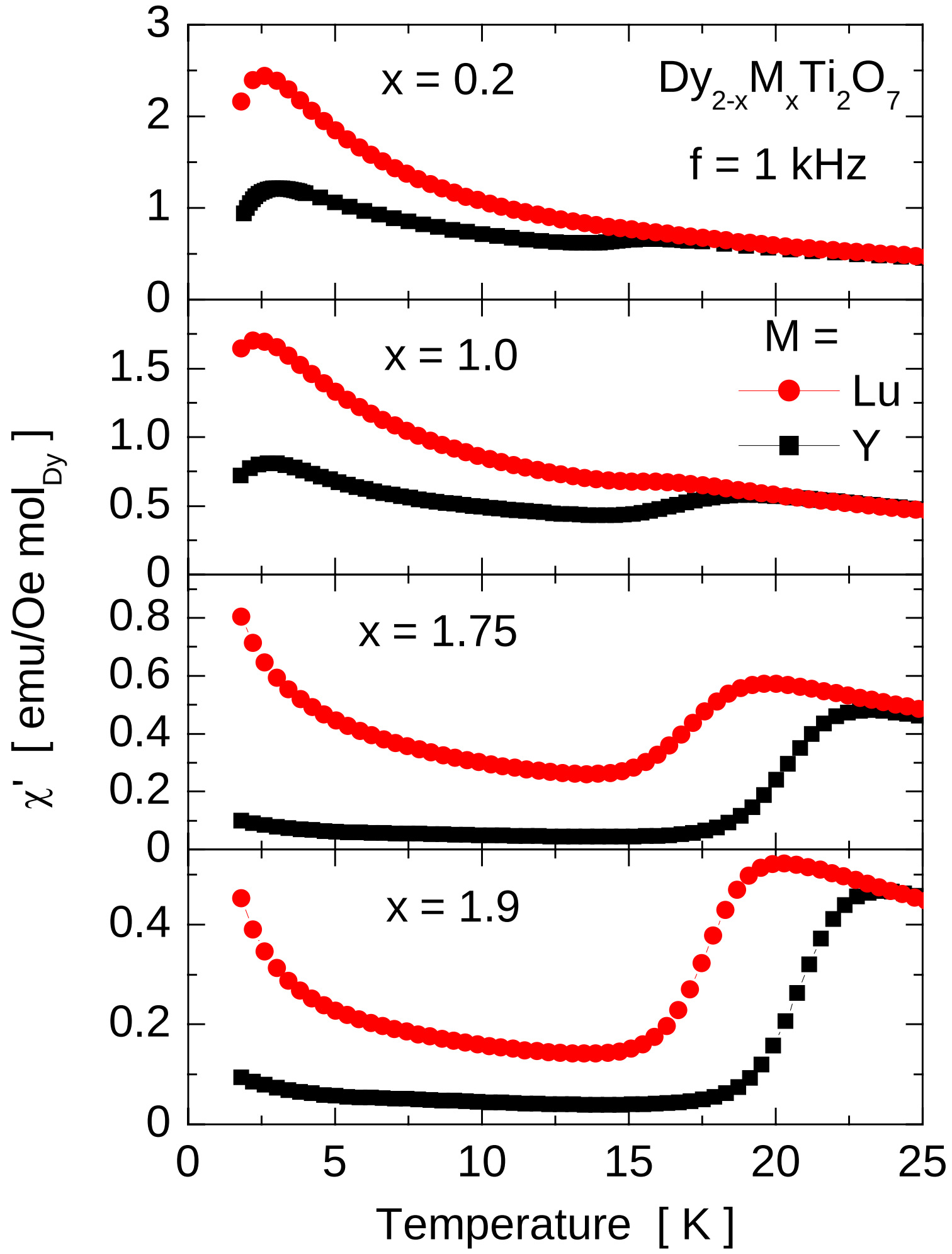

Figure 5 Snyder *et al.*

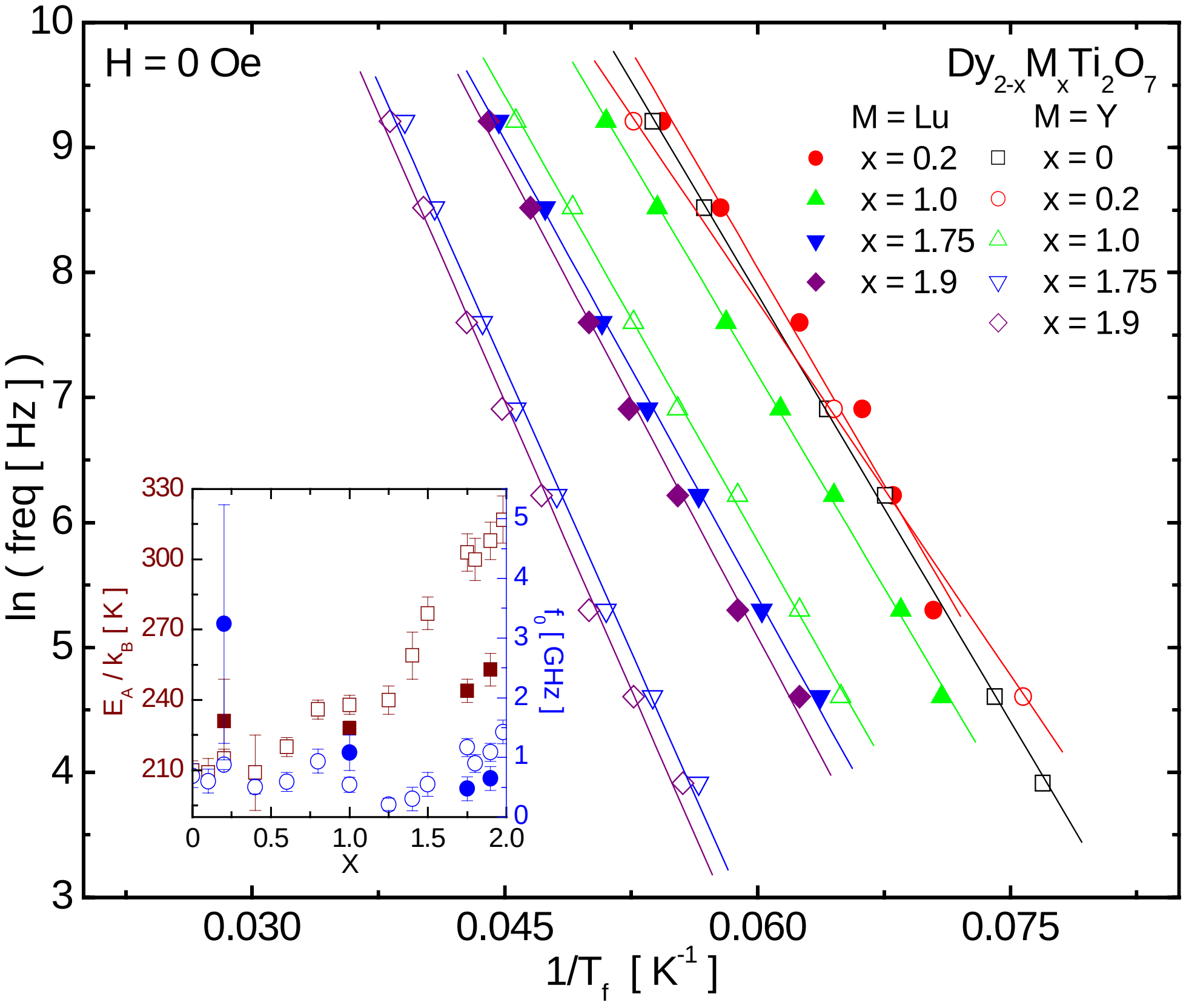

Figure 6 Snyder *et al.*

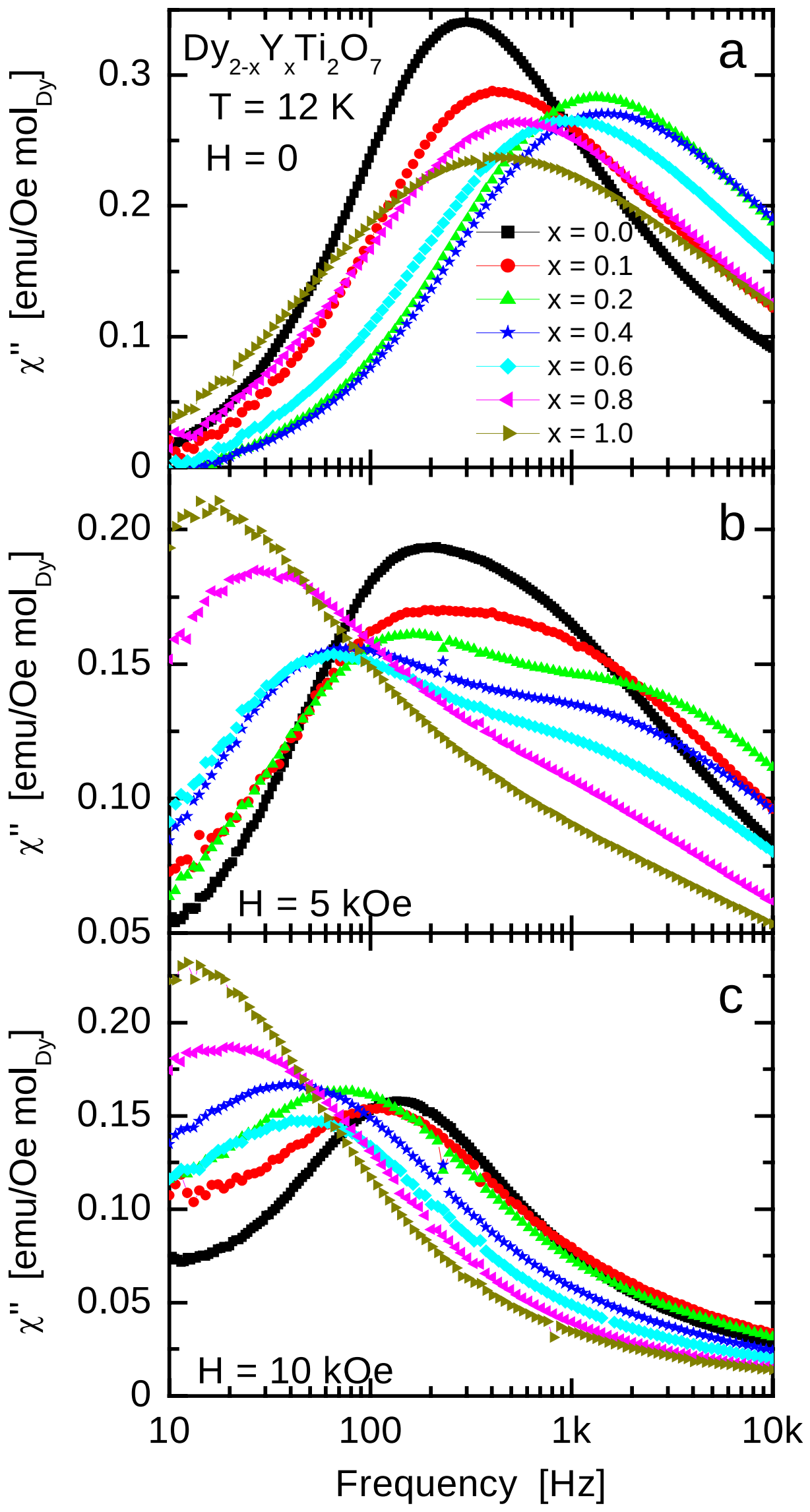

Figure 7   Snyder *et al.*

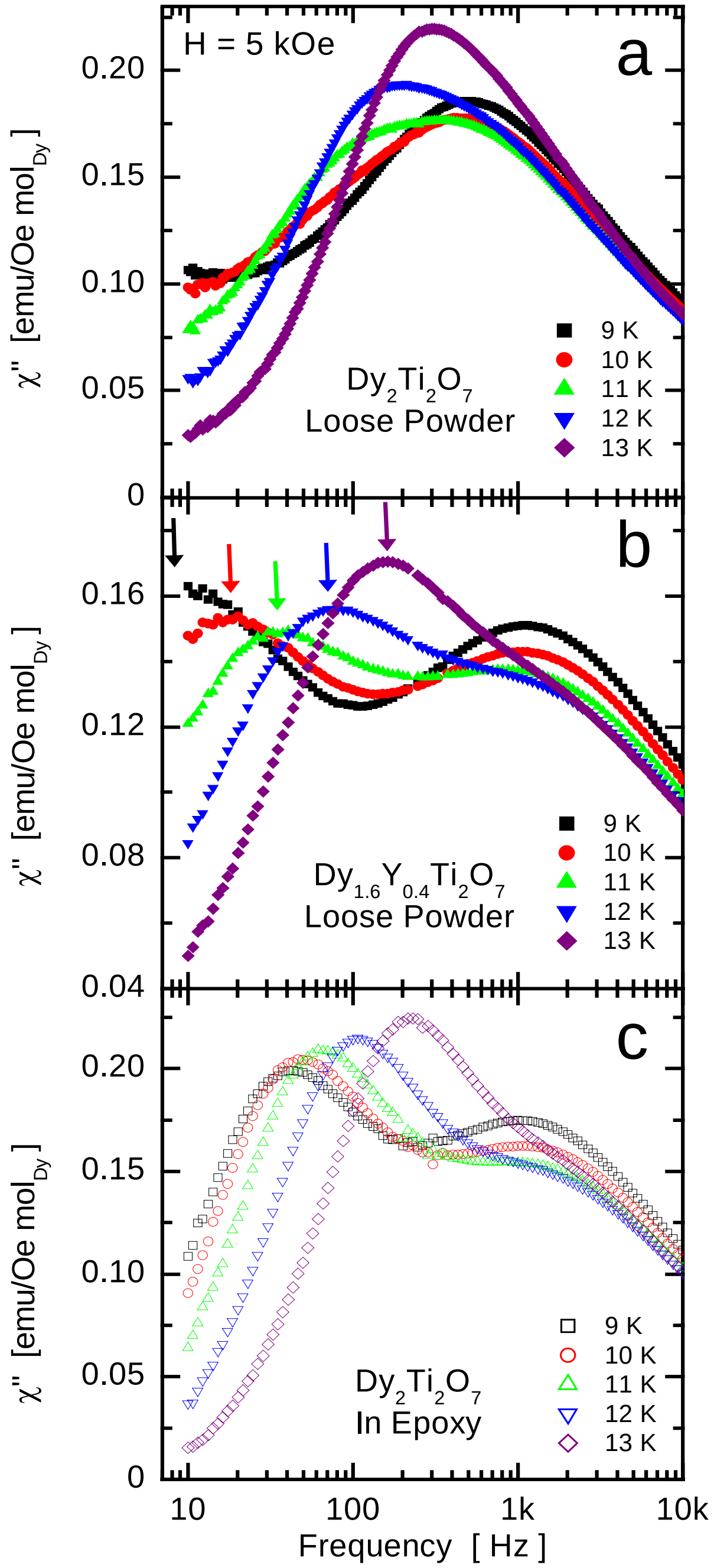

Figure 8 Snyder *et al.*

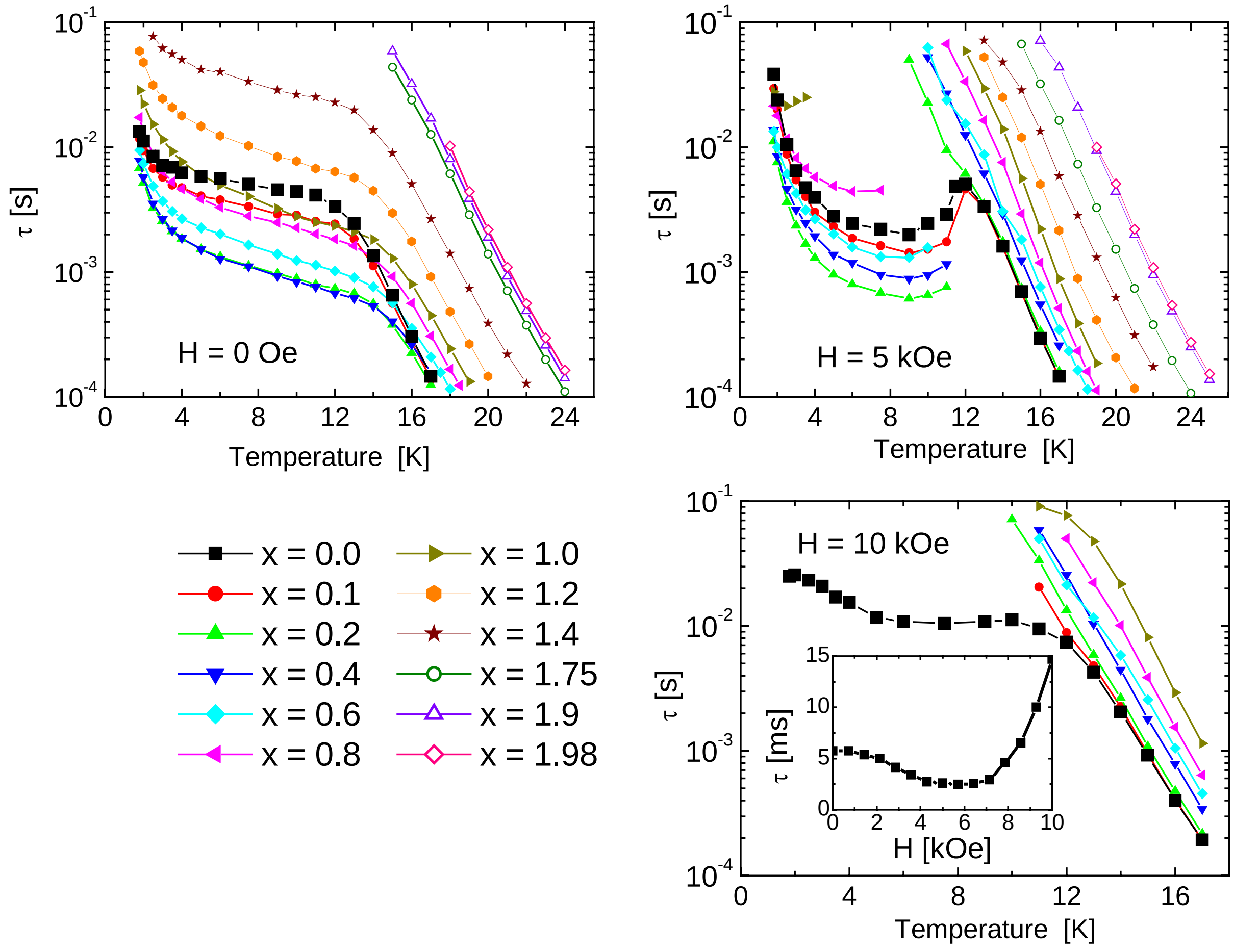

Figure 9   Snyder *et al.*

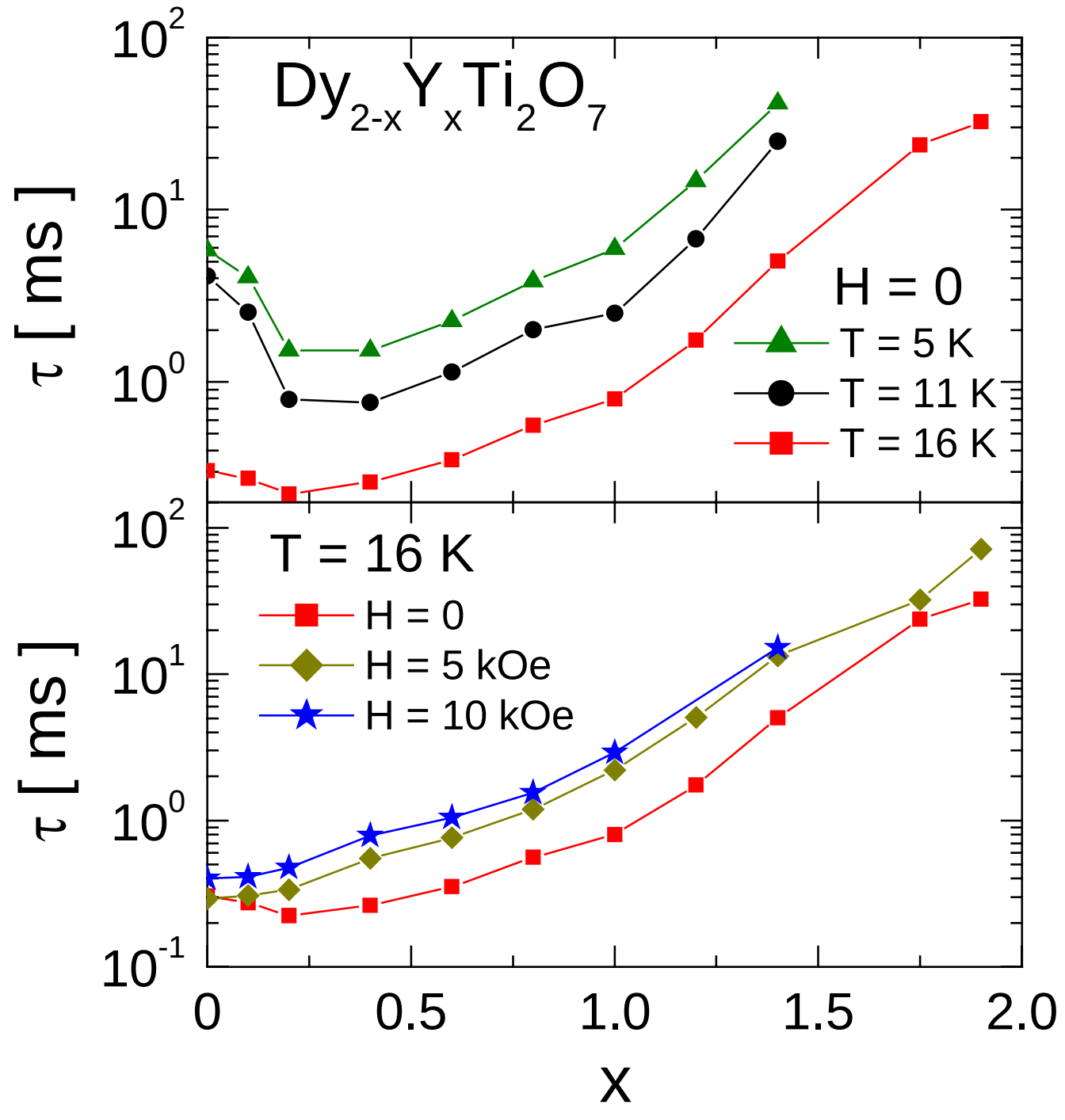